\input harvmac
\newif\ifdraft\draftfalse
\newif\ifinter\interfalse
\ifdraft\draftmode\else\interfalse\fi
\def\journal#1&#2(#3){\unskip, \sl #1\ \bf #2 \rm(19#3) }
\def\andjournal#1&#2(#3){\sl #1~\bf #2 \rm (19#3) }

\def\ie{{\it i.e.}}
\def\eg{{\it e.g.}}

\def\frac#1#2{{#1\over#2}}

\def\half{\frac12}

\def\d{\partial}

\def\inbar{\,\vrule height1.5ex width.4pt depth0pt}
\def\IC{\relax\hbox{$\inbar\kern-.3em{\rm C}$}}
\def\IR{\relax{\rm I\kern-.18em R}}
\def\IP{\relax{\rm I\kern-.18em P}}

%
%
\def\np#1#2#3{Nucl. Phys. {\bf B#1} (#2) #3}
\def\pl#1#2#3{Phys. Lett. {\bf #1B} (#2) #3}
\def\plb#1#2#3{Phys. Lett. {\bf #1B} (#2) #3}

\def\mpl#1#2#3{Mod. Phys. Lett. {\bf #1} (#2) #3}

\catcode`\@=11
\def\slash#1{\mathord{\mathpalette\c@ncel{#1}}}
\overfullrule=0pt

\def\MM{{\cal M}}
\def\NN{{\cal N}}

\def\underrel#1\over#2{\mathrel{\mathop{\kern\z@#1}\limits_{#2}}}

\catcode`\@=12


%

\def\mod{{\rm mod}}

\def\exp{{\rm exp}}


\def\ul{{U(1)}}
\def\nul{{\NN/\ul}}

\def\[{[}
\def\]{]}

\def\comment#1{ }

%
\def\draftnote#1{\ifdraft{\baselineskip2ex
                 \vbox{\kern1em\hrule\hbox{\vrule\kern1em\vbox{\kern1ex
                 \noindent \underbar{NOTE}: #1
             \vskip1ex}\kern1em\vrule}\hrule}}\fi}
\def\internote#1{\ifinter{\baselineskip2ex
                 \vbox{\kern1em\hrule\hbox{\vrule\kern1em\vbox{\kern1ex
                 \noindent \underbar{Internal Note}: #1
             \vskip1ex}\kern1em\vrule}\hrule}}\fi}

%
%

\def\gm{\gamma}                
                
\def\ep{\epsilon}

\def\lm{\lambda}               



               \def\Om{\Omega}
               

%
%
\def\inbar{\hskip.3em\vrule height1.5ex width.4pt depth0pt}
\def\IC{\relax{\inbar\kern-.3em{\rm C}}}
\def\IN{\relax{\rm I\kern-.16em N}}
\def\IQ{\relax\hbox{$\inbar$\kern-.3em{\rm Q}}}
\def\IZ{\relax{\rm Z\kern-.8em Z}}
%
%

\def\Nc{\hbox{$\cal N$}}

\def\hs{\hskip 5mm}
\def\hsc{\hskip 2mm ,\hskip 5mm}
\def\inv{^{-1}}
\def\pt{\partial}
\def\goto{\rightarrow}

%
\rightline{RI-4-99}
\rightline{EFI-99-31}
\Title{
\rightline{hep-th/9907178}}
{\vbox{\centerline{Holography for Non-Critical Superstrings}}}
\medskip
\centerline{Amit Giveon}
\smallskip
\centerline{\it Racah Institute of Physics, The Hebrew University}
\centerline{\it Jerusalem 91904, Israel}
\centerline{giveon@vms.huji.ac.il}
\bigskip
\centerline {David Kutasov, Oskar Pelc}
\centerline{\it Department of Physics, University of Chicago}
\centerline{\it 5640 S. Ellis Av., Chicago, IL 60637, USA }
\centerline{kutasov, oskar@theory.uchicago.edu}
\bigskip\bigskip\bigskip
\noindent
We argue that a class of ``non-critical superstring'' vacua
is holographically related
to the (non-gravitational) theory obtained by studying string theory on a
singular Calabi-Yau manifold in the decoupling limit $g_s\to 0$.
In two dimensions, adding fundamental strings at the singularity of the
CY manifold leads to conformal field theories dual to a recently constructed
class of $AdS_3$ vacua. In four dimensions, special cases of the construction
correspond to the theory on an $NS5$-brane wrapped around a Riemann surface.
\vfill

\Date{7/99}
\newsec{Introduction}

\lref\abks{O. Aharony, M. Berkooz, D. Kutasov and N. Seiberg,
hep-th/9808149, JHEP {\bf 9810} (1998) 004.}

In \abks\ it was proposed that vacua of string theory which
asymptote at weak coupling to linear dilaton backgrounds are
holographic -- string theory in such vacua is dual
to a theory without gravity in a lower dimension. The dual
theory is in general not a local QFT.

\lref\chs{C. Callan, J. Harvey and A. Strominger, hep-th/9112030,
in Trieste 1991, Proceedings, String Theory and Quantum Gravity 1991,
208.}
\lref\natistr{N. Seiberg, hep-th/9705221, \pl{408}{1997}{98}.}
\lref\dvv{R. Dijkgraaf, E. Verlinde and H. Verlinde, hep-th/9604055,
Nucl. Phys. {\bf B486} (1997) 77; hep-th/9704018, Nucl. Phys. {\bf B506} 
(1997) 121.}
\lref\brs{M. Berkooz, M. Rozali and N. Seiberg, hep-th/9704089, Phys. Lett.
{\bf 408B} (1997) 105.}
\lref\ks{D. Kutasov and N. Seiberg, \pl{251}{1990}{67}; for a review,
see D. Kutasov, hep-th/9110041.}

The only example discussed in detail in \abks\ was
string theory in the near-horizon
geometry of $NS5$-branes \chs,
which was argued to be dual to the non-local theory
that governs the dynamics of $NS5$-branes
at vanishing string coupling \natistr\ (see also
\refs{\dvv,\brs}). It was further noted
in \abks\ that a rich set of linear dilaton vacua is
provided by the ``non-critical superstring'' construction
of \ks. The problem of identifying the corresponding non-gravitational
spacetime theory was (in general) left open in \abks.

The main purpose of this note is to fill this gap and propose
dual descriptions for a large class of non-critical superstring models
in $d$ spacetime dimensions\foot{We will consider the cases
$d=2,4,6$.}. We will argue that the dual theory
is obtained by studying string theory on
\eqn\aa{\IR^{d-1,1}\times X^{2n} \hsc 2n=10-d~,}
where $X^{2n}$ is a singular Calabi-Yau manifold.
Sending $g_s\to 0$ at fixed $l_s$ gives rise as in \natistr\
to a $d$ dimensional theory without gravity describing the
dynamics of modes living near the singularity on $X^{2n}$.
This theory is dual to string theory in a background which
approaches at weak coupling $\IR^{d-1,1}\times \IR_\phi\times\Nc$,
where $\IR_\phi$ is the real line along which the dilaton
changes linearly, and $\Nc$ is related to $X^{2n}$
in a way described below.

\lref\adscft{For a review, see O. Aharony, S.S. Gubser, J. Maldacena,
H. Ooguri, Y. Oz, hep-th/9905111.}
\lref\gr{A. Giveon and M. Ro\v{c}ek,
hep-th/9904024, JHEP {\bf 9904} (1999) 019.}
\lref\bl{D. Berenstein and R.G. Leigh,
hep-th/9904040.}
\lref\gks{A.\ Giveon, D.\ Kutasov, and N.\ Seiberg,
hep-th/9806194,
Adv. Theor. Math. Phys. {\bf 2} (1998) 733.}%

In two dimensions, the construction involves string
theory on $\IR^{1,1}\times X^8$, where $X^8$ is a singular CY
fourfold. Adding fundamental strings at the singularity in
$X^8$ and flowing to the infrared leads to a
two dimensional CFT which is dual via the AdS/CFT correspondence \adscft\
to an $N=2$ supersymmetric $AdS_3$ vacuum of the sort recently
discussed in \refs{\gr,\bl} following \gks.

\lref\wita{E. Witten, hep-th/9703166, Nucl. Phys. {\bf B500} (1997) 3.}
\lref\revGK{For a review, see A. Giveon and D. Kutasov, hep-th/9802067.}
In four dimensions, special cases of the construction correspond
to the theory on an $NS5$-brane with worldvolume
$\IR^{3,1}\times\Sigma$, where $\Sigma$ is a Riemann surface;
these examples might be of interest for describing QCD via branes
\refs{\wita,\revGK}.

The plan of this paper is the following. We start in section 2
with a review of the non-critical superstring construction of
\ks. We discuss the symmetry structure of these string vacua and
construct some observables which belong to short representations
of spacetime supersymmetry.

\lref\ginsmoore{For a review, see P. Ginsparg and G. Moore, hep-th/9304011.}

In section 3 we define the dual theories \aa\ and study some
of their properties. We propose the duality and perform a few
simple checks of its validity. We also point out the similarity
of our proposal to the ``duality'' between large $N$ matrix quantum
mechanics in the double scaling limit and $1+1$ dimensional
string theory in a linear dilaton background \ginsmoore.

In section 4 we discuss in turn
two, four and six dimensional examples of our construction. We
make contact with recent work on string theory on $AdS_3$, and
with theories on branes which are relevant for solving strongly
coupled gauge theories. We also discuss the resolution of strong
coupling singularities of non-critical superstrings.

Some of the technical details are contained in an appendix.

\newsec{Non-critical superstrings}

Consider superstring propagation on
\eqn\rdp{\IR^{d-1,1}\times \IR_\phi\times \NN~,}
where $\NN$ is a manifold whose properties will be specified below.
The real line $\IR_\phi$ is parametrized by $\phi$. The dilaton
$\Phi$ is linear in $\phi$,
\eqn\lindil{\Phi=-{Q\over2}\phi~.}
The string coupling $g_s\sim e^\Phi$ goes to zero
as $\phi\to\infty$ and diverges as $\phi\to-\infty$. In
regions where $g_s$ is large, the perturbative
definition of the theory is not useful. To fully
define string theory in the background \rdp, \lindil, one must
either eliminate the strong coupling region, or provide a
definition of the theory at strong coupling.

In the rest of this section we summarize some properties of
the worldsheet description of the vacuum \rdp; this description
is reliable in regions where the string coupling is small.
The linear dilaton \lindil\ implies that the worldsheet
stress-tensor and central charge of $\phi$ are
\eqn\stten{\eqalign{
T_\phi(z)=&-\half\left[(\partial_z\phi)^2+ Q\partial_z^2\phi\right]~,\cr
c_\phi=&1+3Q^2~.\cr}}
The worldsheet superpartner of $\phi$ is $\psi^\phi$; the
superconformal current is
\eqn\tf{T_F=\psi^\phi\partial_z\phi+Q\partial_z\psi^\phi~.}
Consistency of fermionic string propagation requires
the worldsheet theory on $\NN$ to be an $N=1$ SCFT with
central charge $c_{\NN}=3(n-\frac{1}{2}-Q^2)$, where $n$
is given in \aa. We will
also assume that $\NN$ is compact and non-singular, so that
the worldsheet SCFT on $\NN$ is unitary and has a discrete
spectrum of scaling dimensions.

\lref\fms{D. Friedan, E. Martinec and S. Shenker,
Nucl. Phys. {\bf B271} (1986) 93.}

The construction of \ks\ requires that the
worldsheet SCFT on $\NN$ have the following
additional properties:
\item{(a)} An affine $U(1)$ symmetry
with supercurrent \eqn\ucur{\psi^\ul+\theta J^\ul~.}
\item{(b)} The coset $\NN/U(1)$ must have an $N=2$ superconformal
symmetry with central charge
\eqn\cnn{c^{\NN/U(1)}=3(n-1-Q^2)~.}

\noindent
If both of these conditions hold, one can construct a type II string
vacuum with (at least) $2^{{d\over2}+1}$ supercharges as follows.
It is convenient to write the affine current
\ucur\ as
\eqn\bul{J^\ul=i\d Y~,}
where $Y$ is a canonically normalized scalar: $Y(z)Y(w)\sim-\log(z-w)$.
The $N=2$ superconformal algebra on $\NN/U(1)$ contains a $\ul_R$ current
$J^\nul_R$ normalized as
\eqn\rnorm{J^\nul_R(z)J^\nul_R(w) \sim\frac13\,\frac{c^\nul}{(z-w)^2}~,}
which can be expressed
in terms of a canonically normalized scalar $Z$ as
\eqn\br{J^\nul_R=i\sqrt{\frac{c^\nul}3}\d Z\equiv ia\d Z~,\;\;
a\equiv\sqrt{n-1-Q^2}~.}
It is also convenient to bosonize the worldsheet fermions
$\psi^\phi$, $\psi^{U(1)}$ in terms of a
scalar field $H$: $\partial H=\psi^{\phi}\psi^{U(1)}$.
The spacetime supercharges are given by \ks
\eqn\gspace{\eqalign{Q^+_\alpha=&\oint dz\,
e^{-\frac\varphi2}e^{-{i\over2}(H+aZ-QY)}S_\alpha~,\cr
Q^-_{\bar\alpha}=&\oint dz\,
e^{-\frac\varphi2}e^{{i\over2}(H+aZ-QY)}S_{\bar\alpha}~,\cr
}}
where $\varphi$ is the scalar field arising in the bosonized
$\beta,\gamma$ superghost system of the fermionic string,
and $S_\alpha$, $S_{\bar\alpha}$ are spinors of $SO(d-1,1)$ \fms.
For $d=2\; \mod\; 4$, $S_\alpha$, $S_{\bar\alpha}$ are
isomorphic spinors, while for $d=0\; \mod\; 4$
they are distinct.
It is not difficult to check that the supercharges
\gspace\ are BRST invariant and mutually
local on the worldsheet (and thus physical), and that
they form the spacetime superalgebra $(\mu=0,1,\cdots, d-1)$
\eqn\stsupalg{\{Q^+, Q^-\}=\gamma^\mu P_\mu~.}
All the other anticommutators vanish. In \stsupalg\
$P_\mu$ is the momentum along $\IR^{d-1,1}$ and $\gamma^\mu$ are
the corresponding Dirac matrices. The supercharges \gspace\ carry
charge $\pm Q/2$ under the $U(1)$ symmetry \bul, which
is thus an R-symmetry in spacetime. It is customary
to normalize the R-charge so that the supercharges
have $R=\pm1$. Thus we define
\eqn\rcharge{R=i{2\over Q}\oint\partial Y~.}
In a type II string, there are also
supercharges $\bar Q^\pm$ that arise in a similar fashion from the
other worldsheet chirality. For $d=2\;\mod\;4$,
$\bar Q^+$ and $\bar Q^-$ transform
in isomorphic spinor representations of $SO(d-1,1)$. If furthermore
$Q^\pm$ and $\bar Q^\pm$ transform as isomorphic spinors, the theory has chiral
supersymmetry in spacetime; if $Q^\pm$ and $\bar Q^\pm$ transform as
different spinors, the spacetime supersymmetry is non-chiral. This is
analogous to the choice of IIA or IIB strings in ten dimensions.
The full R-charge for the type II case is $R+\bar R$, where $\bar R$
is the antiholomorphic analog of \rcharge.

\lref\seipol{N. Seiberg, Prog. Theor. Phys. Suppl. {\bf 102} (1990) 319;
J. Polchinski, Strings '90, Published in College Station Workshop, 1990.}

Observables in linear dilaton theories correspond to
non-normalizable vertex operators whose wavefunctions
diverge in the weak coupling region $\phi\to\infty$ \seipol. We
next discuss a few examples.
Consider an $N=1$ superconformal primary in the CFT on $\NN$
which is primary with charge $q$ under the $U(1)_R$ symmetry \bul,
and whose projection in $\NN/U(1)$, $V$, is primary under
the full $N=2$ worldsheet superconformal algebra. Such an operator
can be written as
\eqn\opqW{e^{iqY}V~.}
We denote the scaling dimension
of $V$ by $\Delta_V$ and its $U(1)_R$ charge (under \br) by $Q_V$.
Unitarity of the $N=2$ SCFT on $\NN/U(1)$ implies that
\eqn\wunit{\Delta_V\ge {|Q_V|\over2}~,}
with equality when $V$ is a chiral operator in the worldsheet theory
on $\NN/U(1)$.
One can form a physical observable out of \opqW\ as follows:
\eqn\Wdress{e^{-\varphi-\bar\varphi} e^{i\vec k\cdot\vec x+iqY+\beta\phi}V~,}
where $\vec k$ is the momentum along $\IR^{d-1,1}$ and $\beta$ is the
``Liouville dressing.'' The mass-shell condition and GSO
projection (mutual locality of \Wdress\ with \gspace)
lead to the following physical state constraints:
\eqn\physcons{\eqalign{
&\half|\vec k|^2+\half q^2-\half\beta(\beta+Q)+\Delta_V=\half~,\cr
&Q_V-qQ\in 2Z+1~.\cr
}}
Non-normalizability of the wavefunction as $\phi\to\infty$ implies
that \Wdress\ must satisfy\foot{For $\beta=-Q/2$ the non-normalizable
solution is $\phi\exp(-Q\phi/2)$.}
\eqn\boundbeta{\beta\ge -{Q\over2}~.}
One can think of \Wdress\ as a fermionic string ``tachyon''\foot{It is
a tachyon if $V$ is a function on $\NN/U(1)$. In general,
\Wdress\ corresponds to an excited state of the string.};
because of the chiral GSO projection it is of course never tachyonic
\ks.

Another set of observables corresponds to ``gravitons,''
whose $(-1,-1)$  picture vertex operators have the form:
\eqn\dilgrav{e^{-\varphi-\bar\varphi}\xi_{\mu\nu}\psi^\mu \bar\psi^\nu
e^{i\vec k\cdot\vec x+iqY+\beta\phi}V~,}
where $\psi^\mu$ are worldsheet fermions on $\IR^{d-1,1}$,
$\xi_{\mu\nu}$ is the polarization,
and the physical state conditions are\foot{In addition,
there are transversality conditions on $\xi_{\mu\nu}$
which we will not specify.}
\eqn\qkb{\eqalign{
&\half |\vec k|^2+\half q^2-\half \beta(\beta+Q)+\Delta_V=0~,\cr
&Q_V-qQ\in 2Z~.\cr
}}
As in critical string theory, there is
an infinite tower of observables generalizing \Wdress, \dilgrav;
we will not discuss them here.

Chiral operators, that belong to short representations of spacetime
supersymmetry, are of special interest. Such operators can be obtained
by taking $V$ to be a chiral operator on the worldsheet, for which the
inequality \wunit\ is saturated (we will take $Q_V$ to be non-negative,
and $\Delta_V=Q_V/2$), and setting the momentum along $\IR^{d-1,1}$ to zero.
Consider first the ``tachyon'' \Wdress. One can solve the constraints
\physcons\ by setting $\beta=q$, $Q_V-qQ=1$. It can be shown that the
resulting operator belongs to a short representation of spacetime
supersymmetry. The R-charge \rcharge\ of a chiral tachyon operator
\Wdress\ is
\eqn\rV{R_V={2q\over Q}={2(Q_V-1)\over Q^2}~.}
Note that the constraint \boundbeta\ applied to these chiral operators
implies\foot{In the case $\beta=-Q/2$ one does not find a chiral operator
because of the insertion of $\phi$ in front of the exponential (see
footnote 2).} that $Q_V-1>-Q^2/2$ or
\eqn\consnorm{Q_V+{Q^2\over2}-1>0~.}
For $Q^2>2$ this constraint is automatically satisfied,
since by assumption
$Q_V\ge 0$, while for $Q^2<2$ it implies that some of the
chiral operators on the worldsheet do not give rise to chiral
operators in spacetime.

\lref\dixon{L. Dixon, Lectures given at the 1987 ICTP Summer Workshop
in High Energy Physics and Cosmology.}%

The last statement should be qualified somewhat. First,
note that the chiral operators just constructed
are {\it top components} of chiral superfields
in spacetime. One way of seeing that is the following.
In the $(-1,-1)$ picture, the operator
\Wdress\ with $\beta=q$ and $\vec k=0$ has the form
\eqn\chiopp{e^{-\varphi-\bar\varphi} e^{q(\phi+iY)}V~;}
$e^{q(\phi+iY)}V$ is the bottom component of a {\it worldsheet}
chiral superfield with scaling dimension $(1/2,1/2)$.
The $(0,0)$ picture vertex operator is the top component
of this chiral superfield. Such operators can
be added to the worldsheet Lagrangian without breaking
$(2,2)$ superconformal symmetry \dixon, and therefore also
spacetime supersymmetry.

\lref\pp{A.W. Peet and J. Polchinski, hep-th/9809022,
Phys. Rev. {\bf D59} (1999) 065011.}
\lref\ab{O. Aharony and T. Banks, hep-th/9812237,
JHEP {\bf 9903} (1999) 016.}

What is the spacetime interpretation of adding the
operator \chiopp\ to the worldsheet Lagrangian?
According to \abks, the operator \Wdress\
corresponds to an off shell observable $O(\vec x)$ in the
dual, non-gravitational theory\foot{We are speaking
loosely here. The dual theory probably does not have local
observables. It does have observables labeled by arbitrary $d$
dimensional momenta such as \Wdress, \dilgrav,
but there may be subtleties in Fourier
transforming to real space (see \refs{\pp, \ab} for further
discussion).}. The $\vec k=0$ mode of \Wdress, given by
\chiopp, corresponds to $\int d^d\vec xO(\vec x)$, and adding it to
the worldsheet Lagrangian corresponds in the dual theory
to adding $\int d^d \vec x O(\vec x)$ to the spacetime action.
The fact that doing that does not break supersymmetry
implies that $O(\vec x)$ is the top component of a chiral
superfield.

The bottom component of this superfield, $V_{\rm bottom}$,
is killed by half of the supercharges in \gspace, $Q_\alpha^+$.
Acting on $V_{\rm bottom}$ with the remaining
half of the supercharges, the $2^{{d\over2}-1}$ $Q_{\bar\alpha}^-$,
gives rise to \chiopp. Hence the R-charge of $V_{\rm bottom}$ is
\eqn\rVbottom{R_{\rm bottom}={2(Q_V-1)\over Q^2}+2^{{d\over2}-1}~.}
Note that it always satisfies $R_{\rm bottom}\ge {2(Q_V-1)\over Q^2}+1$.
The constraint \consnorm\ implies that $R_{\rm bottom}$ is always positive.

Therefore, what we have found before is that chiral worldsheet
operators which do not satisfy \consnorm\ do not give
rise to observables of the form \Wdress\ which are
{\it top} components of {\it chiral} spacetime superfields.
Instead, they give rise to {\it bottom} components of
{\it antichiral} spacetime superfields. Indeed,
returning to \Wdress, another solution to the physical state
conditions \physcons\ with $\vec k=0$ is $\beta=-q-Q$, $q=(Q_V-1)/Q$.
Since $q$ has the same value as before, the R-charge of
the corresponding spacetime operator,
\eqn\antichiopp{e^{-\varphi-\bar\varphi} e^{-(q+Q)\phi+iqY}V~,}
is again given by \rV.
The condition \boundbeta\ for non-normalizability of \antichiopp\ is
\eqn\consnormbott{Q_V+{Q^2\over2}-1<0~,}
the opposite of \consnorm. The R-charge \rV\ is negative definite
in the regime \consnormbott, which is simply the statement
that the corresponding operator is the bottom component of
an antichiral spacetime superfield.

The top components of spacetime superfields \chiopp, \antichiopp\
play an important role in the theory. As discussed in \refs{\ks,\abks},
due to the linear dilaton \lindil\ the string coupling diverges as
$\phi\to -\infty$. There are different known mechanisms for regulating
this strong coupling divergence. One that was used in \ks, and will play
a role below, is to add to the worldsheet Lagrangian
an operator of the form \chiopp\ with $-Q/2<q<0$.
This sets a scale for fluctuations of $\phi$. For $q<0$
the vertex operator \chiopp\ grows as $\phi\to-\infty$
(the wavefunction, which differs from the vertex operator
by a factor of $g_s$, is always supported at $\phi\to\infty$).
Thus, it prevents the system from running to the strong coupling
region $\phi\to-\infty$. Vertex operators which grow
as $\phi\to-\infty$ can be thought of as relevant operators in
linear dilaton backgrounds. Ref.
\ks\ used the operator \chiopp\ with $V=1$, the ``$N=2$
cosmological constant'' of $N=2$ Liouville theory, to set
the scale. This operator exists only for $Q^2>2$ \consnorm.
For $Q^2<2$ one may try to use a relevant
worldsheet chiral operator with $1>Q_V>1-{Q^2\over2}$,
if one exists.

Alternatively, one may use the top component of an antichiral
superfield associated with \antichiopp. However, in this case
there is the following subtlety. One nice property of \chiopp\
is that if the operator $V$ is the bottom component of a
relevant superfield in the worldsheet SCFT
on $\NN/U(1)$ (\ie\ $Q_V<1$), the dressed operator \chiopp\ is
relevant in spacetime (\ie\ $q<0$), and vice versa. For the
antichiral operators \antichiopp, the relation is more complicated.
For such operators
\eqn\betabot{\beta=-q-Q={1-Q_V-Q^2\over Q}~.}
They are relevant in spacetime (\ie\ have $\beta<0$) if
\eqn\relanti{1-{Q^2\over2}>Q_V>1-Q^2}
(the first inequality is \consnormbott).
As mentioned above, for $Q^2>2$ \relanti\ has no solutions
with $Q_V\ge 0$. For $2>Q^2>1$ all operators \antichiopp\
which satisfy \consnormbott\ are relevant in spacetime
(as well as on the worldsheet). For $Q^2<1$, operators with
$Q_V<1-Q^2$ are relevant on the worldsheet but irrelevant in
spacetime.

\noindent
Comments:
\item{(1)} Our discussion of the relevance in spacetime
concerned the operators \antichiopp, which are bottom
components of spacetime superfields. Since the spacetime
supercharges do not change the $\phi$ dependence of the
wavefunctions, the same analysis holds for the top components,
which are the operators one would actually add to the
worldsheet Lagrangian.
\item{(2)} Equation \betabot\ has the strange property that the
more relevant the operator $V$ is on the worldsheet, the less
relevant the corresponding operator \antichiopp\ is in spacetime.
We do not understand this behavior.
\item{(3)} A similar analysis can be performed for gravitons \dilgrav.
Chiral  operators correspond to $\beta=q=Q_V/Q$. The corresponding
R-charge is
\eqn\rVV{R_V={2Q_V\over Q^2}~.}

\noindent
To recapitulate, the procedure outlined above leads to
a non-critical superstring theory in $d$ dimensions
with $2^{{d\over2}+1}$ supercharges. The theory is
not conformal, and the arguments of \abks\ suggest
that it is holographically related to a (perhaps non-local)
$d$ dimensional theory without gravity.

In the next section we will propose a candidate
for the theory without gravity which is related by the duality
of \abks\ to the string vacua described above. We will
specialize\foot{It is probably possible to generalize our
discussion to the most general compact
manifold $\NN$ satisfying the constraints described above.
This is left for future work.}
to a class of backgrounds \rdp\ for which the
worldsheet CFT on $\NN$ is a product of the $S^1$
\ucur\ and a Landau-Ginzburg (LG) $N=2$ SCFT of $n+1$
chiral superfields $z_a$, $a=1,\cdots,n+1$, with superpotential
\eqn\suppotza{W(z_a)=F(z_a)~,}
where $F$ is a quasi-homogenous polynomial with weight one under
$z_a\to\lambda^{r_a}z_a$, \ie\
\eqn\homog{F(\lambda^{r_a}z_a)=\lambda F(z_a)~, \;\;\lambda\in \IC~,}
for some set of positive weights $r_a$. Here and below we take $F$
to be transverse, \ie\ the only point at which all derivatives
$\partial_{z_a} F$ vanish is the origin, $z_a=0$.

In applications, the worldsheet CFT on $\NN$
is in general not a {\it direct}
product of $S^1$ and the LG model \suppotza. Often, one can reach a
point in moduli space where the two are decoupled, and we will assume
that we are at such a point in the analysis below. It is easy to
generalize to situations where this is not the case.

The worldsheet central charge $c_W$ corresponding to \suppotza\ is
\eqn\cW{{1\over3}c_W=\sum_{a=1}^{n+1}(1-2r_a)=n+1-2\sum_a r_a~.}
It is useful to define
\eqn\rom{r_\Omega\equiv\sum_{a=1}^{n+1} r_a-1~,}
in terms of which
\eqn\cWW{c_W=3(n-1-2r_\Omega)~.}
Comparing \cWW\ to \cnn\ we see that
\eqn\Qrom{Q^2=2r_\Omega~.}
In particular, $r_\Omega$ must be positive in this construction.

Chiral operators are constructed as in the general discussion above.
The worldsheet chiral operators $V$ are in this case
polynomials\foot{$A_i$ are arbitrary quasi-homogenous polynomials,
defined modulo the equations of motion, $\partial_{z_a}F(z_a)=0$.}
in $z_a$, $A_i(z_a)$, which have weights $r_i$ under $z_a\to\lambda^{r_a}z_a$.
Their worldsheet $R$-charges $Q_V$ are equal to $r_i$.
As discussed above, one can construct chiral operators
in spacetime out of the $A_i$ by dressing them in different ways.
The dressing \Wdress\ gives rise to the chiral operator
\eqn\chitach{e^{-\varphi-\bar\varphi} e^{q(\phi+iY)}A_i~,}
where $q=(r_i-1)/Q$; the R-charge \rV\ is
\eqn\Ri{R_i={r_i-1\over r_\Omega}~.}
As in the general discussion, for $r_\Omega<1$ not all homogenous
polynomials $A_i$ give rise by using \chitach\
to top components of chiral superfields in the spacetime theory.
The constraint \consnorm\ implies that only those with
\eqn\consai{r_i+r_\Omega-1>0}
give rise to such operators. The rest of the $A_i$ give bottom
components of antichiral superfields.

Similarly, \dilgrav\ gives rise to a chiral operator
with R-charge $r_i/r_\Omega$ (see \rVV).

\newsec{The dual theory}

\subsec{The dynamics near a singularity}

\lref\OV{H. Ooguri and C. Vafa, hep-th/9511164, \np{463}{1996}{55}.}
\lref\Kutasov{D. Kutasov, hep-th/9512145, \plb{383}{1996}{48}.}
\lref\ghm{R. Gregory, J. Harvey and G. Moore, hep-th/9708086,
Adv. Theor. Math. Phys. {\bf 1} (1997) 283.}%

Consider type II string theory on the manifold $\IR^{d-1,1}\times \bar X^{2n}$,
where $\bar X^{2n}$ is a CY $n$-fold.  If $\bar X^{2n}$ is smooth,
the theory becomes free in the limit $g_s\to 0$. If however $\bar X^{2n}$
contains an isolated singular point, $y_0$, one expects in general to find
in this limit a theory with a scale $l_s$ describing modes localized in the
vicinity of $y_0$. To study the dynamics near the singularity, it
is sufficient to consider the part of the original compact manifold which
is close to $y_0$, and to replace $\bar X^{2n}$ by an appropriate
non-compact manifold $X^{2n}$ with the same singularity. The non-compactness
of $X^{2n}$ also allows one to consider singularities that cannot be embedded
in a compact CY manifold $\bar X^{2n}$.

\lref\kkl{E. Kiritsis, C. Kounnas and D. Lust, hep-th/9308124,
Int. J. Mod. Phys. {\bf A9} (1994) 1361.}%

A well-known example with $d=6$ is obtained by taking $\bar X^{2n}$ to be a
$K3$ manifold with an ADE singularity. In the vicinity of the singular point,
the $K3$ can be replaced by an appropriate
non-compact ALE space.
In the limit $g_s\to 0$ bulk string theory decouples, and one is left with a
non-local theory without gravity with a scale $l_s$ \natistr\foot{In fact,
\natistr\ considered the decoupling limit $g_s\to 0$ in string vacua containing
coincident $NS5$-branes, which are related to ALE spaces by T-duality
\refs{\OV,\Kutasov,\ghm} (see also \kkl).}. 
While only ADE singularities of some fixed finite
order can be embedded in a compact $K3$, the non-compactness of an ALE space
allows one to consider all ADE singularities.

\lref\gvw{S. Gukov, C. Vafa and E. Witten, hep-th/9906070.}

We next list a few properties of these decoupled theories.
We focus on the special case of quasi-homogenous hypersurface
singularities (see \gvw\ for a recent discussion), although it
is probably possible to generalize the discussion to other cases.
Thus, we take the non-compact CY manifold $X^{2n}$ describing
the vicinity of the singularity to be the hypersurface
$F(z_1, \cdots, z_{n+1})=0$ in $\IC^{n+1}$, where $F$ is
a polynomial which transforms as $F\goto\lm F$ under
\eqn\ICtrans{z_a\goto\lm^{r_a}z_a~,}
as in \homog. The point $z_a=0$ is a singular point in $X^{2n}$
and one expects to find a decoupled theory living in its vicinity in
the limit $g_s\to 0$. Isometries of $X^{2n}$ give rise to symmetries
of this theory. In particular, the hypersurface $F(z_a)=0$ has
a $U(1)$ symmetry which acts as \ICtrans\ with $|\lambda|=1$.
The holomorphic $n$-form $\Omega$ on $X^{2n}$,
\eqn\OmDef{\Om=\frac{dz_1\wedge\ldots\wedge dz_n}{\pt F/\pt z_{n+1}}\hsc}
has charge $r_\Om=\sum_a r_a-1$ under this symmetry.
If $r_\Omega\not=0$, this $U(1)$ symmetry
is an R-symmetry. Indeed, since one can write $\Omega$ in terms
of a covariantly constant spinor $\eta$ on $X^{2n}$ as $\Omega_{i_1\cdots
i_n}=\eta^t\Gamma_{i_1\cdots i_n}\eta$, the R-charge of the spacetime
supercharges is $\pm1/2$ that of $\Omega$, \ie\ $\pm r_\Omega/2$.
Moreover, for any hypersurface
singularity which occurs at a finite distance in CY moduli space,
$r_\Omega$ must be positive \gvw. Since we are interested primarily
in singularities that occur at a finite distance, we will restrict to
the case $r_\Omega>0$ below.

The singular hypersurface $F(z_a)=0$ can be deformed to
\eqn\fdef{F(z_a)+\sum_it_i A_i(z_a)=0~,}
where $A_i(z_a)$ are quasi-homogenous polynomials with weight
$r_i$. In the decoupled theory near the singularity, $t_i$
correspond to couplings of top components of chiral superfields.
{}From \fdef\ we see that the $U(1)_R$ charge of $t_i$ is $1-r_i$.
The corresponding operator in the spacetime
theory, which we will also
denote by $A_i$, thus has charge $r_i-1$.
If we normalize the R-charge so that the supercharges have charge
$\pm1$, the charge of $A_i$ becomes
\eqn\rai{{2(r_i-1)\over r_\Omega}~.}
Not all couplings $t_i$ can be turned on in this theory. By requiring
that the kinetic energy for $t_i$ diverge as $z_a\to\infty$, so
that $t_i$ correspond to non-fluctuating couplings, \gvw\ found that
only modes that satisfy \consai\ exist in this theory.

\subsec{The proposed duality}

\lref\phases{E. Witten, hep-th/9301042, \np{403}{1993}{159}; see also
lectures by E. Witten in {\it Quantum fields and strings: a course
for mathematicians}, Am. Math. Soc., 1999, ed. P. Deligne et al.}

We propose that the non-gravitational theory describing
the dynamics of string theory
on $\IR^{d-1,1}\times X^{2n}$ in the limit $g_s\to 0$, where
$X^{2n}$ is a non-compact CY $n$-fold with an isolated
singular point $y_0$, is dual to string theory
on $\IR^{d-1,1}\times\IR_\phi\times\NN$,
where $\NN$ is the manifold consisting of all points
in $X^{2n}$ at a fixed distance from the singular point
$y_0$\foot{More precisely, as we will see below, the string
background involves the worldsheet CFT to which
the sigma model on $\NN$ flows in the IR.}.

For the case of hypersurface singularities $F(z_a)=0$
discussed above, the manifold $\NN$ can be thought of
as $\NN=X^{2n}/\IR_+$ where $\IR_+$ acts on $z_a$ as \ICtrans\
with $\lambda\in \IR_+$. This quotient has a residual $U(1)$
action \ICtrans\ with $|\lambda|=1$ which is the $U(1)_R$ symmetry
discussed above. $\NN/U(1)\simeq X^{2n}/\IC$ is the $n-1$ complex dimensional
hypersurface $F(z_a)=0$ in the $n$ dimensional weighted projective space
$W\IC P^n_{r_1,\ldots,r_{n+1}}$.

Is the manifold
$\NN$ described above a good background for string
propagation? As discussed in section 2, for this to be
the case the non-linear sigma-model on $\NN/U(1)$ must
be $(2,2)$ superconformal. One way to think about this
issue is to embed this non-linear sigma model in a gauged
linear sigma-model following \phases. One studies a $(2,2)$ supersymmetric
$U(1)$ gauge theory with matter fields $z_a$ ($a=1,\cdots, n+1$)
with charges $r_a$ and an additional field $P$ with charge $-1$.
The gauge invariant superpotential is taken to be
\eqn\gaugesup{W=PF(z_a)~.}
Since the gauge group is Abelian, one can add to the Lagrangian
a Fayet-Iliopoulos D-term, $r$. Classically, when $r$ is large
and positive the low energy dynamics of the linear sigma model
corresponds to the non-linear sigma-model on $\NN/U(1)$, and $r$
can be thought of as the size of the space (a K\"ahler modulus).
When $r$ is large and negative one finds a LG model\foot{More precisely, 
this is an orbifold of the above LG model, which here is implemented
by the GSO projection.} with $W=F$ \phases.

Quantum mechanically, the behavior of
the theory depends on the sum of the gauge charges of the matter
fields \phases. When this sum vanishes, the quantum picture is closely
related to the classical one: $r$ remains a modulus in the
quantum theory (\ie\ it is truly marginal), and changing
this modulus between $-\infty$ and $\infty$ interpolates
between LG theory and the non-linear sigma-model. Thus,
in this case both are
conformal, and provide different descriptions of a single
moduli space of CFT's.

When the sum of the charges is non-zero, there are
logarithmic corrections to $r$ at one loop. Thus,
there is a $\beta$-function for $r$ and it flows
as we change the scale. When the sum of the charges
is positive, $r$ decreases as we go to longer distances,
and vice-versa. The detailed RG flows in this case
have not been analyzed, to our knowledge.

In our case, the sum of the gauge charges is
$r_\Omega=\sum_a r_a-1$ and, as we saw before,
it is positive. Thus, one expects the K\"ahler modulus
$r$ of the hypersurface $\NN/U(1)$ to decrease
with increasing scale, and presumably go
to $r=-\infty$ at long distances. It is then
natural to expect that the
infrared fixed point of the non-linear sigma-model
on $\NN/U(1)$ is the infrared limit of the LG theory with superpotential
$W=F$ discussed in section 2 \suppotza\ (we will denote it below
by $LG(W=F)$). Therefore, the duality proposed in the beginning
of this subsection involves in this case the background
$\IR^{d-1,1}\times\IR_\phi\times\NN$, where $\NN$ is
roughly $S^1\times LG(W=F)$. As mentioned before, the
product here may not be direct, \ie\ there may be a correlation
between the quantum numbers of operators in $S^1$ and in $LG(W=F)$.

The duality proposed here is very reminiscent of the ``old
matrix model'' \ginsmoore. There, one considers quantum mechanics
of $N\times N$ Hermitian matrices in a potential $V(M)$ in the large
$N$ limit. The Lagrangian is
\eqn\Lmatrix{\CL={\rm Tr}\left(\half \dot{M}^2-V(M)\right)~.}
For generic $V$ the dynamics depends sensitively
on the detailed structure of $V$. However, one can fine tune
the couplings in $V$ and approach a codimension one surface
in coupling space along which the dynamics described by \Lmatrix\
is singular. The physics near this singularity is universal
(\ie\ independent of the detailed structure of $V$), and for the
purpose of studying it one can replace $V$ by an inverted harmonic
oscillator potential.

In the double scaling limit $N$ is taken to infinity while approaching
the singular surface in coupling space, keeping a dimensionful parameter
which measures the distance from the singularity fixed. This limit leads to
a non-trivial theory. Green functions of $U(N)$ invariant observables
have non-trivial $1/N$ expansions and one can attempt to study
non-perturbative effects in $1/N$ as well. As described in \ginsmoore,
these Green functions have an alternative description as S-matrix elements
in a $1+1$ dimensional string theory of a scalar field and
the Liouville field $\phi$. The $1/N$ expansion in the double scaled
matrix model
is equivalent to the genus expansion of this string theory. The dilaton
is linear in $\phi$, and the dimensionful parameter
which measures the distance (in coupling space) of
\Lmatrix\ from the singularity becomes the
worldsheet cosmological constant.

Our construction is very similar. String theory on $\IR^{d-1,1}\times
\bar X^{2n}$ is the analog of the matrix QM \Lmatrix. In particular,
it contains a lot of ``non-universal'' information.
Approaching a point in CY moduli space where $\bar X^{2n}$
develops a singularity is analogous to tuning the couplings
in the potential $V$ to the singular surface.
Replacing $\bar X^{2n}$ by $X^{2n}$
is the analog of replacing a general $V$ near the critical surface
by the inverted harmonic oscillator potential. The limit $g_s\to 0$
in string theory on $\IR^{d-1,1}\times X^{2n}$ is the analog of the limit
$N\to\infty$ in the matrix model. Finally, the non-critical superstring
on $\IR^{d-1,1}\times \IR_\phi\times \NN$ \rdp\ is the analog of
$1+1$ dimensional string theory with a linear dilaton.

The analog of turning on the worldsheet cosmological
constant in our case should be taking the limit $g_s\to 0$ for a slightly
resolved singularity, as in \fdef. Indeed, we will see below that
in some cases one can deform the singularity from $F(z_a)=0$ to
$F(z_a)=\epsilon$, and $\epsilon$ becomes the non-critical superstring
``cosmological constant'' discussed in \ks, which prevents the system
from running to strong coupling.

We next list a few simple checks of the proposed duality:
\item{(1)} The symmetry structure seems to agree. Both models
have $2^{{d\over2}+1}$ supercharges and a $U(1)_R$ symmetry.
\item{(2)} The parameter $r_\Omega$ defined by \rom\ plays
an important role in both models, and must be positive in
both. In string theory on $\IR^{d-1,1}\times\IR_\phi\times S^1
\times LG(W=F)$ this is due to \Qrom, while in
$\IR^{d-1,1}\times X^{2n}$ it is due (for example)
to the requirement that the singularity be at a finite
distance in moduli space of CY manifolds \gvw. In both theories
the physics changes at $r_\Omega=1$ ($Q^2=2$).
\item{(3)} Agreement of the chiral rings: we
saw that deformations $A_i$ of the polynomial $F$
\fdef\ give rise to chiral superfields whose top components
have R-charge $2(r_i-1)/r_\Omega$. In the theory near the
singularity on $X^{2n}$ the R-charge is given by \rai, while in the linear
dilaton vacuum it is eq. \Ri\foot{The R-charge given in \Ri\
appears to be too small by a factor of two. The total R-charge receives
a contribution from the other worldsheet chirality. The operators
\chitach\ are left-right symmetric; hence, the total R-charge found
in section 2 is $2(r_i-1)/r_\Omega$, in agreement with \rai.}.
Furthermore, in both approaches it was found that only deformations
that satisfy \consai\ are allowed. In both descriptions, the origin
of this constraint is the requirement that the corresponding coupling
be non-normalizable as $\phi\to\infty$. 

\newsec{Some special cases}

\subsec{Two dimensional models $(d=2)$}

In this case, the duality proposed in the previous section
relates the following models with four supercharges in
$1+1$ dimensions:
\item{(1)} Type II string theory on
\eqn\doneone{\IR^{1,1}\times \IR_\phi\times S^1\times LG(W=F)~,}
where the LG superpotential is $W=F(z_1,\cdots, z_5)$.
\item{(2)} Type II string theory on $\IR^{1,1}\times X^8$
in the limit $g_s\to 0$, where $X^8$ is the singular
CY fourfold $F(z_1,\cdots, z_5)=0$.

\noindent
For type IIA, models (1) and (2) have $(2,2)$ supersymmetry
in $1+1$ dimensions.
For IIB they have chiral $(4,0)$ supersymmetry.

In some cases, one can think of theory (2) as a theory
on an $NS5$-brane, making the connection to \refs{\natistr,
\abks} more apparent. For example, if we take
\eqn\FADE{F(z_1,\cdots, z_5)=H(z_1,z_2,z_3)+z_4^2+z_5^2~,}
where $H(z_1,z_2,z_3)$ describes an ADE singularity or a deformation
thereof, theory (2) above can be thought of as the decoupled theory
on a curved $NS5$-brane whose worldvolume is $\IR^{1,1}\times L_4$,
where $L_4$ is the hypersurface $H(z_1,z_2,z_3)=0$ in $\IC^3$ \gvw.
This follows\foot{We thank K. Hori and H. Ooguri for pointing this
out to us.} from a straightforward generalization of the arguments
of \OV.
Because of the usual chirality flip between fivebranes and singular
geometries (which is essentially due to T-duality), theories (1)
and (2) in type IIA are related to $NS5$-branes in IIB, and vice versa.

\lref\kehag{A. Kehagias, hep-th/9805131, Phys. Lett. {\bf B435}
(1998) 337.}
\lref\klwit{I.R. Klebanov and E. Witten, hep-th/9807080,
Nucl. Phys. {\bf B536} (1998) 199.}
\lref\afhs{ B.S. Acharya, J.M. Figueroa-O'Farrill, C.M. Hull
and B. Spence, hep-th/9808014, Adv. Theor. Math. Phys. {\bf 2} (1998) 1249.}
\lref\morpl{D.R. Morrison, M.R. Plesser, hep-th/9810201.}

In theory (2), one can add $p$ fundamental
strings at the singular point in $X^8$, $z_a=0$.
In type IIA this does not break any further supersymmetry,
while in IIB it breaks $(4,0)$ supersymmetry to $(2,0)$.
In the infrared, the resulting theory will generically approach
a non-trivial $(2,2)$ or $(2,0)$ superconformal fixed point. It
is natural to expect that the description of this fixed point in
the dual theory (1) is obtained by replacing $\IR^{1,1}\times
\IR_\phi$ in \doneone\ by $AdS_3$, and studying string theory on
\eqn\adsone{AdS_3\times S^1\times LG~.}
This seems to be consistent with recent work on the AdS/CFT correspondence
for branes at singularities \refs{\kehag,\klwit,\afhs,\morpl,\adscft}.

\lref\oog{J. de Boer, H. Ooguri, H. Robins and J. Tannenhauser,
hep-th/9812046, JHEP {\bf 9812} (1998) 026.}%
\lref\ksp{D.\ Kutasov and N.\ Seiberg, hep-th/9903219,
JHEP {\bf 9904} (1999) 008.}%
The class of vacua \adsone\ was recently discussed in \refs{\gr,\bl} who
showed that the spacetime theory indeed has $N=2$ superconformal symmetry.
The spacetime central charge of these vacua is \refs{\gks,\oog,\ksp}
\eqn\cspacetime{c_{\rm spacetime}=6kp~,}
where $k$ is the radius of curvature of $AdS_3$, or equivalently
the level of the $SL(2)$ current algebra on the worldsheet of
the string, and $p$ the number of strings. Equation \cspacetime\
is valid for all $k$ and to leading order in $1/p$.

\lref\sw{N. Seiberg and E. Witten, hep-th/9903224,
JHEP {\bf 9904} (1999) 017.}
By comparing the worldsheet central charge of the linear dilaton
vacuum \stten\ to that of $AdS_3$, one finds that $k$ is related
to the parameters $Q$, $r_\Omega$ as follows:
\eqn\kqrom{{1\over k}={Q^2\over2}=r_\Omega~.}
In sections 2,3 we saw that the physics of
non-critical superstrings depends
on whether $r_\Omega$ is larger or smaller
than one. It is interesting to note that string theory
on $AdS_3$ also undergoes a kind of phase transition
at the point $k=r_\Omega=1$. As discussed in \sw\ (see also \ksp),
string theory on $AdS_3$ has a set of excitations corresponding
to long strings living at the boundary of $AdS_3$. The energy
gap for the system to emit one of these strings is finite, and
it goes to zero as $k\to 1$. Excitations of long strings form
a continuum above the gap. A related fact is that these long
strings become ``critical'' at $k=1$. For $k>1$ their string
coupling grows as one approaches the boundary of $AdS_3$, while
for $k<1$ it goes to zero there \sw.

It would be interesting to understand better the relation between
the phase transitions observed in the linear dilaton, singular CY
and $AdS_3$ systems.

The analysis of excitations performed for the linear dilaton
vacuum in section 2 can be repeated for $AdS_3$. For example,
the analogs of the observables \Wdress\ in this case are
\eqn\adsW{e^{-\varphi-\bar\varphi} e^{iqY}V\Phi_h~,}
where $\Phi_h$ is a primary of $SL(2)$ in the spin $j=h-1$
representation; its spacetime scaling dimension is $h$ (see
\ksp\ for a more detailed discussion). The physical state
constraints in this case are
\eqn\adsphyscons{\eqalign{
&\half q^2-{h(h-1)\over k}+\Delta_V=\half~,\cr
&Q_V-qQ\in 2Z+1~.\cr
}}
Chiral operators under the spacetime $N=2$ superconformal
algebra have scaling dimension $h$ equal to one half their
$R$-charge\foot{As shown in \refs{\gr,\bl}, \rcharge\ is the zero mode
of the spacetime $U(1)$ current that is part of the spacetime
$N=2$ superconformal algebra.} \rcharge. Thus we set
\eqn\chircond{h={|q|\over Q}=|q|\sqrt{k\over2}~.}
Plugging in \adsphyscons\ we find
\eqn\cchh{{h\over k}+\Delta_V={1\over2}~.}
Taking $V$ to be a chiral operator with
$\Delta_V=Q_V/2$ as before, we find that
\eqn\spchdim{h={k\over2}(1-Q_V)~.}
The second equation in \adsphyscons\ together with \chircond\ then implies
that $q$ is negative, and $h=-q/Q$. Thus, the operator we found is a
bottom component of an antichiral superfield. In string theory
on $AdS_3$ only operators $\Phi_h$ with $h>1/2$ exist\foot{The
scaling dimension of $\Phi_h$ and $\Phi_{1-h}$ is $-h(h-1)/k$.
$\Phi_h$ with $h>1/2$ is physical,
while the other solution is related to it by an integral
transform \ref\teschn{J. Teschner, hep-th/9712256; hep-th/9712258.},
whose convergence requires that $h>1/2$. An alternative way to
see that only $h>1/2$ gives rise to non-normalizable observables
is to study the behavior of the wavefunction $\Phi_h$ at the boundary
of $AdS_3$, as done in \ksp.}. Imposing this
in \spchdim\ leads to the constraint on charges
\eqn\conschrg{Q_V+{1\over k}-1<0~,}
which is nothing but the constraint \consnormbott\ found for the
corresponding operators in the linear dilaton background.

To find analogs of \chiopp\ in $AdS_3$, we require that the
bottom component of the spacetime superfield whose top component
corresponds to \adsW\ be chiral; this implies
\eqn\imph{h-{1\over 2}={1\over 2}\left({2q\over Q}+1\right) \Rightarrow
h-1={q\over Q}~.}
Plugging in \adsphyscons\ leads to
\eqn\hhhh{h=1+{k\over2}(Q_V-1)~.}
This is the dimension of the top component of the superfield.
The dimension of the bottom component (which corresponds to a
RR vertex operator) is
\eqn\hhhhbot{h_{\rm bottom}={1\over2}+{k\over2}(Q_V-1)~,}
and the constraint that in \hhhh\ $h>1/2$, or equivalently
that $h_{\rm bottom}>0$,
\eqn\conschrgchi{Q_V+{1\over k}-1>0~,}
is the same as the constraint \consnorm\ satisfied by
\chiopp.

Therefore, we see that there is a nice correspondence between
the spectrum of operators in the non-conformal linear
dilaton background \doneone, and its conformal low energy
limit \adsone. In fact, as we discuss in the appendix, one
can construct string backgrounds which interpolate between
a linear dilaton vacuum for $\phi\to\infty$ and $AdS_3$ for
$\phi\to-\infty$, which makes this correspondence natural.

We conclude this subsection with an example. Take
\eqn\exF{F(z_1,\cdots, z_5)=z_1^n+z_2^2+z_3^2+z_4^2+z_5^2~,}
which corresponds to an $A_{n-1}$ singularity in \FADE.
The weights $r_a$ are: $r_1=1/n$, $r_2=r_3=r_4=r_5=1/2$.
Hence, $r_\Omega$ \rom\ is given by
\eqn\romex{r_\Omega=1+{1\over n}={n+1\over n}~.}
Note that in this case $r_\Omega$ is always bigger than one,
so the subtleties
of $r_\Omega<1$ discussed in sections 2,3 are never encountered.
The spacetime central charge \cspacetime\ is
\eqn\csex{c_{\rm spacetime}=6(1-{1\over n+1})p~.}
The chiral operators $A_i$ \fdef\ corresponding to
resolutions of the singularity are
$A_i=z_1^i$, $i=0,1,2,\cdots, n-2$. Their $U(1)_R$
charges are $Q_{A_i}=i/n$. Plugging into \hhhhbot\ we find
chiral operators with spacetime scaling dimensions
\eqn\spscddd{h_i={i+1\over 2(n+1)}~,\qquad i=0,1,\cdots, n-2~.}
This spectrum looks like that corresponding to
an $N=2$ minimal model with superpotential $\Phi^{n+1}$.
The central charge \csex\ is also suggestive of that.
Indeed, if we could trust \csex\ for $p=1$ (which is
far from obvious), it would be natural to expect that
in the spacetime SCFT with central charge
\eqn\cssseee{c_{\rm spacetime}=6(1-{1\over n+1})=
\left[3-{6\over n+1}\right]+3~,}
the contribution in square brackets comes from
an $N=2$ minimal model with the above superpotential.
For large $p$, \csex\ might describe a (deformation of a) symmetric
product of such minimal models, each coupled to a $c=3$ system.
It would be interesting to understand to what extent
this is true, and what is the role of the $c=3$ piece
of the spacetime CFT.

A system closely related to ours was recently
discussed in \gvw, who however set $p$ to zero and instead
turned on some discrete RR backgrounds that are difficult to
analyze in our formalism. Interestingly, in the simplest case
it was found in \gvw\ that the spacetime SCFT corresponding
to \exF\ is an $N=2$ minimal model with superpotential $\Phi^{n+1}$
and the result \spscddd\ for the dimensions of chiral operators
was obtained as well. This is encouraging, since
our analysis is reliable at large $p$, while that of
\gvw\ works for small $p$; it would be interesting
to understand the relation between the two.

Returning to the linear dilaton vacuum with $p=0$, the example
\exF\ allows us to demonstrate another point that was briefly
mentioned above. In the non-critical superstring construction,
the string coupling diverges as $\phi\to-\infty$. In the theory
of the $NS5$-brane discussed in \abks\ the strong coupling
singularity is avoided in a way that cannot be understood
in string perturbation theory. Thus, the linear dilaton description
has limited utility; \eg\ it cannot be used for computing correlation
functions.

In our case, the situation is the following. If we are studying
the singular theory, \ie\ take $X^8$ to be the manifold $F(z_1,\cdots,
z_5)=0$, where $F$ is given by \exF, the dual description again
involves an infinite throat with infinite coupling down the throat,
and perturbatively nothing stops the theory from running to $\phi\to
-\infty$. The singularity is again avoided non-perturbatively.
However, we can resolve the singularity
$F=0$ to $F=\epsilon$, by turning on the operator
$A_0=1$ in \fdef. In the non-critical superstring this corresponds
to adding to the worldsheet Lagrangian the operator \chitach\ with $A_i=1$.
This is precisely the operator used in \ks\ to cut off the strong coupling
singularity at $\phi\to-\infty$. After the resolution, string perturbation
theory is well defined, and can be used reliably to calculate
correlation functions.

As discussed in section 3, all this is very
reminiscent of the ``old matrix model'' \ginsmoore. The case of the unresolved
singularity $F=0$ corresponds to vanishing cosmological constant in
Liouville theory (or, equivalently, vanishing condensate of the ``tachyon''
field in $1+1$ dimensional string theory), which is a singular limit,
at least perturbatively. The resolved system $F=\epsilon$ corresponds
to finite cosmological constant and much of the intuition developed
in $1+1$ dimensional string theory
is applicable here\foot{Of course, unlike \ginsmoore, one does not
expect the full string theory in non-critical
superstring vacua to be exactly solvable.}.

Note also that for $p\not=0$, using the result \spscddd\
for the dimensions of chiral operators, we see that $A_0=1$
corresponds to an operator with scaling dimension $h_0={1\over 2(n+1)}$
in the low energy spacetime CFT. Thus, the spacetime superpotential
$\Phi^{n+1}$ is deformed for finite $\epsilon$ to $W=\Phi^{n+1}+
\epsilon\Phi$, which indeed completely resolves the $n$-fold singularity at
$\Phi=0$. This is consistent with the fact that the non-critical
superstring with finite $N=2$ cosmological constant does not
appear to describe a non-trivial spacetime CFT in the infrared.

\lref\kut{D. Kutasov, hep-th/9207064, Mod. Phys. Lett. {\bf A7} (1992) 2943;
E. Hsu and D. Kutasov, hep-th/9212023, Nucl. Phys. {\bf B396} (1993) 693.}

One can also study RG flows in the boundary theory, by adding relevant
deformations to $F$, as in \fdef. This leads to a picture like that
of \kut. The general relevant perturbation of the worldsheet
superpotential \exF\ is
\eqn\gensup{F(z_1,\cdots, z_5)=z_1^n+\epsilon+
\sum_{i=1}^{n-2}\lambda_i z_1^i
+z_2^2+z_3^2+z_4^2+z_5^2~;}
$\epsilon$ sets the scale, and as in \ginsmoore\ it can be scaled to one.
The $\lambda_i$ can then be thought of as dimensionful
couplings that ``flow with the scale.'' The flow can be considered
either as a function of $\epsilon$ or as a function of $\phi$. For
large $\phi$, or equivalently large $\epsilon$, the $\lambda_i$ are
effectively small and the theory approaches the one with $\lambda_i=0$.
For $\phi\to-\infty$, or equivalently $\epsilon\to 0$, the $\lambda_i$
grow and the system generically splits into a set of decoupled vacua with
$n=2$ in \exF. By tuning the $\lambda_i$ one can reach a large collection
of multicritical points (see \kut\ for a more detailed discussion).

\subsec{Four dimensional models $(d=4)$}

\lref\conifold{P. Candelas, P. Green and T. Hubsch,
\np{330}{1990}{49}; P. Candelas and X. de la Ossa, \np{342}{1990}{246}.}
\lref\ArDou{P.C. Argyres and M.R. Douglas,
hep-th/9505062, Nucl.Phys. {\bf B448} (1995) 93.}
\lref\APSW{P.C. Argyres, M.R. Plesser, N. Seiberg and E. Witten,
hep-th/9511154, Nucl. Phys. {\bf B461} (1996) 71.}
\lref\EHIY{T. Eguchi, K. Hori, K. Ito and S-K. Yang,
hep-th/9603002, Nucl. Phys. {\bf B471} (1996) 430.}

The duality of section 3 relates in this case the following
$N=2$ supersymmetric theories in $3+1$ dimensions:
type II string theory on
\eqn\dthreeone{\IR^{3,1}\times \IR_\phi\times S^1\times LG(W=
F(z_1,\cdots, z_4))}
and type II string theory on
\eqn\dualdth{\IR^{3,1}\times X^6}
in the limit $g_s\to 0$. $X^6$ is the singular
CY manifold $F(z_1,\cdots, z_4)=0$.

A simple example is
\eqn\conif{F=z_1^2+z_2^2+z_3^2+z_4^2~,}
for which $X^6$ is the conifold \conifold. $r_\Omega=1$ in this case,
and the LG factor in \dthreeone\ degenerates.

\lref\klmvw{A. Klemm, W. Lerche, P. Mayr, C. Vafa and N. Warner,
hep-th/9604034, Nucl. Phys. {\bf B477} (1996) 746.}%

More generally, if
\eqn\fourADE{F(z_1,\cdots, z_4)=H(z_1,z_2)+z_3^2+z_4^2~,}
we can think of the theory on \dualdth\ as the worldvolume
theory on an $NS5$-brane wrapped around the Riemann surface
$H(z_1, z_2)=0$ \klmvw. Such fivebranes are relevant for describing
$N=2$ SYM theories using branes \refs{\wita, \revGK}.

For example, to study $NS5$-branes wrapped around the Seiberg-Witten
curve at an Argyres-Douglas
point \refs{\ArDou,\APSW,\EHIY}, one can choose $H$ in \fourADE\
to describe an ADE singularity
\eqn\HADE{H(z_1,z_2)=\cases{
z_1^n+z_2^2&$A_{n-1}$\cr
z_1^n+z_1z_2^2&$D_{n+1}$\cr
z_1^3+z_2^4&$E_6$\cr
z_1^3+z_1z_2^3&$E_7$\cr
z_1^3+z_2^5&$E_8$\cr
}}
For $d=4$, $n=3$ \aa, and plugging in \cWW\ gives $c_W=6(1-r_\Omega)$.
Since $c_W\geq 0$, this implies that $r_\Omega\leq 1$. Therefore,
in these examples one typically
encounters the obstructions discussed in sections
2,3 to turning on various perturbations that resolve the
singularities.

For the $A_{n-1}$ singularity \HADE\ $r_\Omega={1\over n}+{1\over2}$.
Perturbations of the form $z_1^i$ ($i=0,1,\cdots, n-2$),
with $r_i={i\over n}$, give rise to deformations of the singularity
only when \consai\ holds, \ie\ for
\eqn\inminone{i>{n\over2}-1~.}
Therefore, an $A_{n-1}$ singularity cannot be completely resolved.

A possible explanation of \inminone\ in the fivebrane
theory is the following. At low energy, the theory of the $NS5$-brane
wrapped around the Riemann surface $H(z_1,z_2)=0$
flows to a four dimensional $N=2$ SCFT
\refs{\ArDou,\APSW,\EHIY}. Such theories have a global $U(1)_R
\times SU(2)_R$ symmetry. Dimensions of chiral operators are
related to the $U(1)_R$ charge $R$ and $SU(2)_R$ spin $I$ via
\eqn\dirr{D=2I+{R\over2}~.}
If the $U(1)_R$ symmetry which becomes part
of the $N=2$ superconformal algebra in the IR can
be identified at high energies, one
can use \dirr\ to determine the dimensions of chiral operators.

In our case, the high energy theory has a $U(1)_R$ symmetry
$R+\bar R$ \rcharge, and it is natural to expect this symmetry
to become the $U(1)_R$ part of the global symmetry of the IR
SCFT. The deformations $z_1^i$ discussed above have $I=0$, hence
their scaling dimensions in the low energy SCFT are $D_i=
{1\over2}(R_i+\bar R_i)=R_i$. Using \rVbottom\
(with $Q_{z_1^i}=r_i={i\over n}$ and $Q^2=2r_\Omega={n+2\over n}$)
we find that the dimension
of the bottom component of the superfield $z_1^i$ is
\eqn\dinmt{D_i=2{i+2\over n+2}~,\qquad i=0,1,\cdots, n-2~.}
All scaling dimensions in a unitary four dimensional
CFT satisfy $D\ge 1$. Imposing this requirement on \dinmt\ leads
to \inminone\ (for $i={n\over 2}-1$, $D_i=1$ and the corresponding
operator is a decoupled free field).

As explained in section 2, operators $z_1^i$ which do not satisfy
\inminone\ give rise via \antichiopp\ to bottom components of antichiral
superfields. Since $Q^2=2r_\Omega$ satisfies $1<Q^2<2$, these
superfields are relevant. Their top components
can be added to the worldsheet
Lagrangian to eliminate the strong coupling singularity, however,
they do not correspond to new deformations of $H$ \HADE.
One can show\foot{More generally, for $d=4$ one can show that
the worldsheet chiral ring of $LG(W=F)$ splits into two
components of equal size: operators that satisfy \consai\
and those that do not. The two groups are related by spectral flow. 
This implies that the size of the spacetime chiral ring is
half of what one might naively expect.}
that the top component of the antichiral superfield
whose bottom component is \antichiopp\ with $V=z_1^i$ is
the complex conjugate of \chiopp\ with $V=z_1^{n-2-i}$.

A similar analysis can be performed for the D, E series curves in \HADE.
Equation \dinmt\ takes in general the form
\eqn\dingenADE{D_i=2{e_i+1\over h+2}~,}
where $h$ is the dual Coxeter number of the corresponding
algebra and $e_i$ are its Dynkin exponents \refs{\ArDou,\APSW,\EHIY}.

\subsec{Six dimensional models $(d=6)$}

In this case, the duality of section 3 relates type II string theory on
\eqn\aga{\IR^{5,1}\times \IR_\phi\times S^1\times LG(W=F(z_1,z_2,z_3))}
and type II string theory on
\eqn\agb{\IR^{5,1}\times X^4}
in the limit $g_s\to 0$. $X^4$ is the singular manifold $F(z_1,z_2,z_3)=0$.
The models \aga\ and \agb\ have sixteen real supercharges. In the type IIA
theory they form a non-chiral $(1,1)$ supersymmetry algebra in $5+1$
dimensions, while in type IIB one finds chiral $(2,0)$ supersymmetry.

The worldsheet central charge of the LG model in \aga\ is \cWW\
\eqn\agbb{c_W=3-6r_\Omega~.}
For positive $r_\Omega$ this central charge is smaller than three.
All unitary $N=2$ SCFT's with central charge $c<3$ have been classified.
They correspond to $N=2$ minimal models and are in one to one
correspondence with ADE singularities. One way of describing them
is as infrared fixed points of LG models with superpotential
\eqn\agc{F(z_1,z_2,z_3)=H(z_1,z_2)+z_3^2~,}
where $H(z_1,z_2)$ is given in eq. \HADE. The manifold $X^4$
\agb\ $F(z_1,z_2,z_3)=0$ is an ALE space corresponding to the appropriate
ADE singularity.

In this case
our proposal reduces to that of \abks, where it was argued that
the decoupled theory on $\IR^{5,1}\times {\rm ALE}$ is holographically
related to string theory on
\eqn\agd{\IR^{5,1}\times \IR_\phi\times S^3~.}
At first sight this appears to be different from the background
\aga, but the two are in fact related as follows.

\lref\gko{P. Goddard, A. Kent and D. Olive, Phys. Lett.{\bf 152B} (1985) 88.}
\lref\gpr{For a review, see A. Giveon, M. Porrati and E.
Rabinovici, hep-th/9401139, Phys. Rept. {\bf 244} (1994) 77.}%

The worldsheet CFT on $S^3$ is described by an $SU(2)$ WZW model
associated with the ADE singularity corresponding to $X^4$. The
$N=2$ minimal models $LG(W=F)$ in \aga\
can be described as the coset SCFT's $SU(2)/U(1)$.
One can decompose $SU(2)$ WZW theory under $U(1)\times SU(2)/U(1)$ \gko.
The GSO projection fixes the radius of the $U(1)$ and
acts as an orbifold on $U(1)\times SU(2)/U(1)$;
this orbifold is equivalent by T duality
to $SU(2)$ WZW CFT \gpr, in agreement with \agd.

As an example, for the $A_{n-1}$ singularity $F(z_1,z_2,z_3)=z_1^n
+z_2^2+z_3^2$, $r_\Omega={1\over n}$. None of the relevant
perturbations $z_1^i$ ($i=0,1,\cdots, n-2$) satisfy \consai.
Hence, they give rise to bottom components of antichiral
superfields. They are killed by the eight supercharges
$Q_\alpha^-, \bar Q_\alpha^-$.
The top components of the superfields are obtained
by acting on them with the eight remaining supercharges
$Q_\alpha^+$, $\bar Q_\alpha^+$.

The perturbations $z_1^i$ are relevant on the worldsheet
but none of them are relevant in spacetime. Indeed, since
$Q^2=2r_\Omega=2/n$, the condition \relanti\ implies that
$Q_{z_1^i}=i/n>1-(2/n)$, or $i>n-2$, outside the range of
available perturbations. Thus, we recover the conclusion
of \refs{\chs,\abks} that in this case there are no relevant
deformations
that can be turned on that eliminate the strong coupling
singularity at $\phi=-\infty$ perturbatively.

\lref\witcig{E. Witten, Phys. Rev. {\bf D44} (1991) 314.}

It should be mentioned for completeness that there is in fact
a known way to eliminate the strong coupling singularity both
in \aga, and more generally in all vacua of the form $\IR^{d-1,1}
\times \IR_\phi\times S^1\times \NN/U(1)$.
$\IR_\phi\times S^1$ is an infinite cylinder; the dilaton
grows as one goes down the cylinder. One can eliminate
the strong coupling region by changing the topology of the
cylinder to the semi-infinite cigar, which can be described
by CFT on $SL(2)/U(1)$ \witcig.

The string coupling on the
cigar is bounded and, in principle, one should be able to
study the theory using worldsheet methods. Since observables
are exponentially supported far from the tip of the cigar,
where the space looks like a cylinder, much of our discussion
above applies to this geometry. To compute correlation functions
it is important to take into account scattering from the tip
of the cigar. This can be done using results on CFT on Euclidean
$AdS_3$ (the coset $SL(2, \IC)/SU(2)$) and coset CFT techniques.

\lref\sfet{K. Sfetsos,
hep-th/9811167, JHEP {\bf 9901} (1999) 015;
hep-th/9903201.}

For example, string theory on
\eqn\slsu{{SL(2)\over U(1)}\times {SU(2)\over U(1)}\times\IR^{5,1}}
is a vacuum with sixteen supercharges which looks
asymptotically like \aga, however, unlike \aga,
it should be a weakly coupled theory\foot{The symmetry
structure of this vacuum is discussed in
Appendix B of \gks.}. According to
\sfet, \slsu\ is related to rotating $NS5$-branes.

\lref\MGV{S. Mukhi and C. Vafa, hep-th/9301083, Nucl. Phys. {\bf B407}
(1993) 667; D. Ghoshal and C. Vafa, hep-th/9506122, Nucl. Phys. {\bf B453}
(1995) 121.}%
\lref\ggkk{A. Giveon and D. Kutasov, hep-th/9909110.}%

\bigskip
\noindent{\bf Note added:} String propagation in the ``near-horizon''
geometry of CY manifolds with hypersurface singularities was 
also studied in \refs{\MGV,\OV}. The relation of our work 
to these papers is discussed in \ggkk.

\bigskip
\noindent{\bf Acknowledgements:}
We thank K. Hori, N. Seiberg and A. Tseytlin for useful discussions.
We also thank the organizers of the String Workshop at the
Institute for Advanced Studies at the Hebrew University for
hospitality and for creating a stimulating environment during
the initial stage of this work.
The work of A.G. is supported in part by the Israel
Academy of Sciences and Humanities -- Centers
of Excellence Program, and by BSF -- American-Israel Bi-National
Science Foundation. D.K. and O.P. are supported in part by
DOE grant \#DE-FG02-90ER40560.

\appendix{A}{Interpolating between linear dilaton and $AdS_3$ vacua}

In the text we mentioned the fact that the two dimensional linear
dilaton $\IR^{1,1}\times \IR_\phi$ and $AdS_3$ vacua are closely
related. We also mentioned that there are solutions which interpolate
between the two. In this appendix we construct such solutions. While
the construction is general, we present it for the special case of
vacua of the form
\eqn\mthree{\MM_3\times S^3\times T^4~,}
where $\MM_3$ interpolates
between $\IR^{1,1}\times \IR_\phi$ for $\phi\to\infty$ and $AdS_3$
for $\phi\to-\infty$. We will first describe the solution in supergravity,
and then the corresponding exact worldsheet CFT.

\lref\DGHR{A. Dabholkar, G. Gibbons, J. Harvey and F. Ruiz-Ruiz,
\np{340}{1990}{33}.}
\lref\Tseytlin{A. Tseytlin, hep-th/9601177, \mpl{A11}{1996}{689}.}
Consider a configuration of $k$ $NS5$-branes  wrapped on a four-torus
of volume $vl_s^4$,
parametrized by the coordinates $x_i$, $i=1,2,3,4$. In the remaining
non-compact dimensions the fivebranes look like $k$ strings whose
worldsheet is the $(\gamma,\bar\gamma)$ plane. One can add to this
configuration $p$ fundamental strings parallel to the fivebranes
(\ie\ extended in $(\gamma,\bar\gamma)$ as well).
The metric, dilaton and $NS$ $B_{\mu\nu}$
field for this configuration of branes, with the $p$ strings
smeared over the four-torus, are \refs{\chs,\DGHR,\Tseytlin}:
\eqn\geoof{\eqalign{
ds^2 & = f_1\inv l_s^2 d\gm d\bar\gm + f_5(dr^2 + r^2 d\Om_3^2) +
dx_i dx^i~, \cr
e^{2\Phi} & = g_s^2 f_5 f_1\inv~, \cr
dB & = \frac{2ip g_s^2}{v} f_5 f_1\inv *_6 \ep'_3 + 2ik\ep'_3~,}}
where $d\Om_3^2$ and $\ep'_3$ are
the metric and volume form on the unit 3-sphere, $*_6$ is the Hodge dual in
the six dimensions parametrized by $(\gm,\bar\gm,r,\Om_3)$ and
\eqn\ff{
f_j = 1 + \frac{R_j^2}{r^2} \hsc R_1^2 = \frac{p g_s^2 l_s^2}{v}
\hsc R_5^2 = k l_s^2~.}
At weak coupling the fivebranes are much heavier than the strings
and thus they give rise to a larger distortion of the geometry around
them (\ie\ typically $R_1\ll R_5$).
Therefore, it makes sense to study an intermediate region in the
background \geoof\ where one is in the near-horizon geometry of
the fivebranes but not necessarily of the strings. As is clear
from \ff, this is the region $r\ll R_5$. In this limit, the geometry
has the form \mthree, where the three dimensional manifold $\MM_3$
is described by:
\eqn\geoofn{\eqalign{
ds_3^2 & = f_1\inv l_s^2 d\gm d\bar\gm + R_5^2 d\phi^2~, \cr
e^{2\Phi} & = \frac{v}{p} e^{-2\phi} f_1\inv~, \cr
dB & = 2i e^{-2\phi} f_1\inv \ep_3 =
d\left[if_1\inv d\gm\wedge d\bar\gm\right]~,}}
where
\eqn\fone{
f_1 = 1 + \frac{1}{k} e^{-2\phi} \hsc
e^\phi = \frac{l_s r}{R_1 R_5} = \sqrt{\frac{v}{pk}} \frac{r}{g_s l_s}~,}
and $\ep_3=*_6 \ep'_3=\frac{1}{k}f_1\inv d\gm\wedge d\bar\gm\wedge d\phi$
is the volume form defined by $ds_3^2/R_5^2$.
For $R_1 \ll r$,
$f_1\approx1$ \ff\
and $\MM_3$ looks like flat space with a linear dilaton,
$\IR^{1,1}\times \IR_\phi$. For $r\ll R_1$ one is in the
near-horizon geometry of both the strings and the fivebranes
and $\MM_3$ becomes the familiar $AdS_3$ solution:
\eqn\geoads{\eqalign{
ds_3^2 & = kl_s^2(e^{2\phi} d\gm d\bar\gm + d\phi^2)~, \cr
e^{2\Phi} & = \frac{kv}{p}~, \cr
dB & = 2ik \ep_3 = d\left[ike^{2\phi}d\gm\wedge d\bar\gm\right]~.}}
Therefore, \mthree\ interpolates between the linear dilaton and
$AdS_3$ vacua.

While the above discussion took place in supergravity,
one can show that the background \geoofn\ in fact gives rise
to an exact solution of the (classical) string equations of
motion, \ie\ to a worldsheet CFT with the right properties.
This CFT can be constructed as a perturbation of CFT on $AdS_3$.
We next briefly review that construction.

\lref\hs{S.F. Hassan and A. Sen, hep-th/9210121, Nucl. Phys.
{\bf B405} (1993) 143.}
\lref\gkir{A. Giveon and E. Kiritsis, hep-th/9303016,
Nucl. Phys. {\bf B411} (1994) 487.}
\lref\forste{S. Forste, hep-th/9407198, Phys. Lett. {\bf B338} (1994) 36.}

CFT on $AdS_3$ has an $SL(2)\times SL(2)$ current algebra. One
of the null holomorphic currents, $J$, and its (anti-holomorphic)
complex conjugate, $\bar J$, can be written semiclassically as
\eqn\apa{J\sim e^{2\phi}\partial\gamma~, \qquad
\bar J\sim e^{2\phi}\bar\partial\bar\gamma~.}
The operator $J\bar J$ is exactly marginal in CFT on $AdS_3$.
Adding this perturbation to the $AdS_3$ background with a finite
coefficient, along
the lines of \refs{\hs,\gkir}, gives rise to a one parameter family
of sigma models with Lagrangian \forste
\eqn\apb{{1\over 2\pi}\left(k\partial\phi\bar\partial\phi
+{k\alpha\over 1+\alpha e^{-2\phi}}\partial\gamma\bar\partial\bar\gamma\right)
+{1\over 8\pi}\sqrt{g}R^{(2)}
\{\log[e^{-2\phi}(1+\alpha e^{-2\phi})^{-1}]+const\}~.}
For $\alpha=1/k$ the sigma model \apb\ is identical to
the background \geoofn,\fone. Thus, the background $\MM_3$
corresponding to \geoofn\
is an exact CFT with $c=3+6/k$.

\listrefs
\end